\documentclass[1p]{elsarticle}

\usepackage{graphicx}
\usepackage{caption}
\usepackage{subcaption}
\usepackage{xcolor}
\usepackage[fleqn]{amsmath}
\usepackage{hyperref}
\usepackage{float}
\usepackage{dblfloatfix}
\usepackage{epstopdf}

\usepackage{lineno}
\modulolinenumbers[5]

\usepackage{setspace}
\newcommand*{\figref}[1]{Fig.~\ref{#1}}
\graphicspath{ {./EPS/} }

\begin{document}
	
\begin{frontmatter}
\title{True Gaussian shaping for high count rate measurements of pulse amplitudes}
\author[phti]{M.Yu.~Kantor \corref{cor1}}
\ead{m.kantor@mail.ioffe.ru}

\author[phti]{A.V.~Sidorov}
\ead{sidorov@mail.ioffe.ru}	 
\address[phti]{Ioffe Institute, Polytekhnicheckaya 26, Saint-Petersburg, Russia, 194021}

\cortext[cor1]{Corresponding author}

\date{}

\begin{abstract}
A digital shaper for high-count-rate detection and amplitude measurement of pulses is proposed and analysed in this paper. The proposed shaper converts pulses with a short leading edge and a long exponential tail into a true Gaussian form. The width of Gaussian pulses can be several times smaller than the rise time of the input pulses, i.e. considerably shorter than the undistorted output pulses provided by standard shapers. Therefore, the proposed true Gaussian shaper resolves strongly overlapped pulses better and provides a higher output count rate. The capabilities of the proposed true Gaussian shaper are analysed with real and simulated output signals of a silicon drift detector of soft X-ray radiation, operating at a high count rate of the collected quanta. Our analysis shows that true Gaussian shapers can increase the count rate of spectrometer systems several times compared with the widely used trapezoidal shapers, while maintaining their amplitude resolution.

\end{abstract}


\end{frontmatter}

\nolinenumbers
\section{Introduction}
\label{sec:Introduction}

The shaping of output pulses from detectors operating in count mode is widely used to resolve overlapped detector pulses and improve their energy resolution. The shaping of detector signals is primarily realised using digital signal processing algorithms \cite{smith2003digital, proakis2001digital}. There are a number of standard shapers \cite{jordanov1994digital,jordanov1994digital1} which convert detector pulses with a short rise time~$\tau_r$ and a long decaying tail $\tau_t$ to shorter pulses whose full width at half maximum (FWHM) is $T_s$. At $T_s\gg\tau_r$, shapers provide well-formed symmetrical output pulses at a low noise level, which ensure unbiased and accurate measurements of pulse amplitudes at count rates lower than $\approx1/T_s$.

At shorter pulse widths ($T_s$), the form of the shaped pulses is altered, output noises are increased, and the inferred pulse amplitudes are biased from their real values. These factors restrict the output count rate of standard shapers to a value significantly lower than $1/\tau_r$.  

The proposed true Gaussian shaper can shape detector pulses into a Gaussian form with $T_s<\tau_r$ to achieve an output count rate for pulse processing several times higher compared with standard shapers. A shaper is defined by its transmission function, calculated as the ratio of the Fourier transforms of the output true Gaussian pulse and the input detector pulse. The capabilities of true Gaussian shaping were tested in an AXAS-D system \cite{KETEKAXAS} with a silicon drift detector (SDD) H7 VITUS \cite{KETEKVITUS} developed by KETEK GmbH in 2012 for high throughput measurements of soft X-ray spectra. The system will be referred hereafter in this paper as the KETEK spectrometer. We will show that the proposed true Gaussian shapers can achieve the maximal output count rate of the spectrometer system, which is several times higher than that obtained with standard shapers, and maintains the energy resolution of the measurements.

The layout of this paper is as follows. First, the measured impulse response of the KETEK spectrometer is presented in Section 2 prior the analysis of true Gaussian shapers, because the impulse response determines the transmission function of the shaper. Two analytical approximations of the impulse response are introduced and are used further in the analysis. In addition, the experimental noise characteristics of the detector are presented in the same section, and are used further for estimation of the dead time, energy resolution and output count rate of spectrometers with true Gaussian shapers. 

The measured impulse response allows for calculation of the transmission function of true Gaussian shapers, presented in Section 3, for output Gaussian pulses of different widths ($T_s$). The shaper exhibited a very strong suppression of signals at frequencies higher than $1/T_s$ for pulse widths $T_s<<\tau_r$. 

This is the strongly suppressed transmission characteristic at high frequencies that provides the short true Gaussian pulses in the shaper output. Gaussian pulses are compared with the output pulses of trapezoidal shapers in Section 4. Trapezoidal pulses keep a symmetrical and undistorted form when their width $T_s>>\tau_r$. Distortions in these short trapezoidal pulses restrict the output count rate of spectrometer systems.  

The amplification of detector noise in true Gaussian and trapezoidal shapers is analysed in Section 5 while taking into account the measured noise spectrum of the KETEK system. The dead times of true Gaussian and trapezoidal shapers are calculated in Section 6. For large pulse widths, the calculated dead time coincided with the peaking plus flat top times of trapezoidal shapers. The concept of resolving time is introduced in Section 7 to specify the shortest time interval between two overlapped pulses when their inferred and actual amplitudes differ less than the output shaper noise. The dead and resolving times, along with the amplitude measurement errors, are given in the aforementioned section as a function of the width of the shaped pulses. 

The output count rate and energy resolution of the KETEK spectrometer with the true Gaussian and trapezoidal shapers are calculated in Section 8 for a wide range of input count rates. The effects of amplitude variations in the shaped pulses with charge collection times in the SDD sensor are analysed in Section 9 for the KETEK spectrometer and a fast SDD with a larger sensor area. The design of a true trapezoidal shaper is discussed in Section 10. The results of this work are summarised in Section 11. 	

\section{Detector impulse response and noise}
\label{sec:Responses}

A true Gaussian shaper is specified by its transmission function, which is the ratio of the Fourier transforms of the output and input pulses. The proposed shaper should always be designed for the actual impulse response of the employed detector for a single quantum, whereas standard shaper are designed for ideal input pulses with zero rise time \cite{jordanov1994digital,jordanov1994digital1}. Therefore, the proposed shaper is analysed with both actual and ideal input pulses to compare it with standard shapers. The characteristics of the actual detector impulse response and its approximations are considered below. 

An AXAS-D spectrometer \cite{KETEKAXAS} with a silicon drift detector of standard class H7 VITUS \cite{KETEKVITUS} made by KETEK GmbH was selected for testing and verifying the proposed shaper. The sensitive area of the detector was 7~mm$^2$. The system was developed for spectral measurements of soft X-ray radiation at output count rates up to $3\cdot10^5~s^{-1}$ at an energy resolution of ~139~eV at FWHM for a photon energy of 5895~eV. The analogue output of the AXAS-D system was used as the input of the tested shapers. The analogue impulse response on a single quantum was measured while using a $^{55}Fe$ radiation source emitting $\approx10^4$ photons with an energy of 5895~eV per second on the detector area. Approximately 4000~pulses from these photons were digitized by a custom built ADC with a resolution of 12~bits and a sampling rate of 50~MHz. The passband of the digitizer ranged from 0 to 25~MHz with a signal amplitude damping of $\omega^{-3}$ outside the band.

\begin{figure*}[h]
	\centering
		\includegraphics[width=\linewidth]{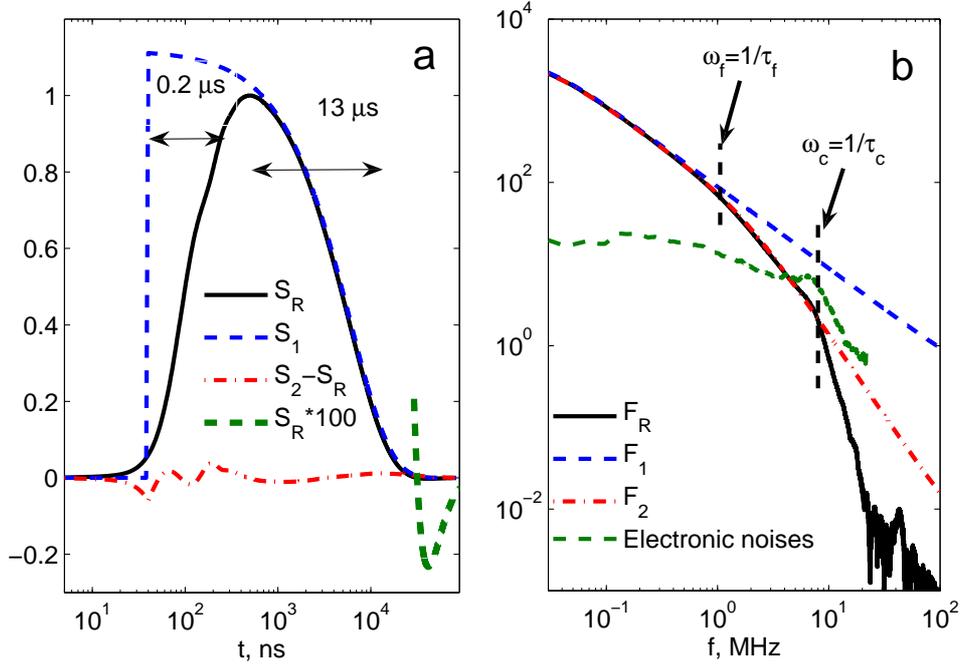}
		\caption{\\a. Impulse response $S_R$ of KETEK spectrometer and its approximations\\b. Fourier spectra of the impulse response, its approximations and detector noises}
		\label{fig:stp}
\end{figure*}

The normalized response of the KETEK system, hereafter referred to as $S_R$ pulse, is shown in \figref{fig:stp}a by a black solid curve. The pulse was interpolated at a sampling period of 2~ns and with an amplitude accuracy of 0.08 \% by averaging the recorded pulses after fine synchronisation of their leading edges. The FWHM of the pulse is 4.6~$\mu$s. The rise time between 0.1 and 0.9 of the pulse amplitude was measured to be 203~ns. The tail of the pulse after scaling by a factor of 100 (plotted in the figure with a dark green dashed curve at $t>10^4$ ns) shows an undershoot of 0.2 \%.

The collection times of the electron clouds created by photon impacts in the H7 absorption layer \cite{Gatti1987, Metzger2004} were estimated to be under 6~ns (see Section 9). They are considerably less than the rise time of $\approx$200~ns of the detector response determined using filters of the AXAS-D preamplifier. Therefore, the rise time is slightly affected by variations in the charge collection time. Estimated experimental variations of the rise time of the impulse response of 5895~eV photons were $\approx$10~ns and were fully accounted for the noise of the KETEK spectrometer.   

Small rise time variations of $\approx$2~ns were measured by the KETEK spectrometer by collecting high-energy 100-keV photons from the high temperature plasma of the FT-2 tokamak ~\cite{lashkul2001effect}. These variations were also accounted for when determining the preamplifier noise. The amplitude variations of the shaped pulses with the charge collection time of the H7 detector and for SDDs with a larger sensor area are analysed in Section 9.   

Two analytical approximations of the impulse response ($S_R$) are also shown in the aforementioned figure. The simplest approximation of the pulse  $S_1=A_1\cdot exp(-t/\tau_t)$ with zero rise time and an exponentially decaying tail, $\tau_t$=5.95~$\mu$s, is shown in \figref{fig:stp}a by a blue dashed curve. The exponential pulse can be converted with standard digital shapers \cite{jordanov1994digital,jordanov1994digital1} to a symmetrical output pulse of trapezoidal, cusp, or pseudo-Gaussian forms. Trapezoidal shapers are known for having the lowest delta noise \cite{Goulding1983} and for their unbiased amplitude measurements of overlapped pulses. Therefore, they are widely used in detection systems, and are thus compared with true Gaussian shapers in the paper in terms of count rate, resolution, and biasing of pulse amplitude measurements.

A more accurate approximation is one that takes into account a finite rise time of the impulse response, but assumes instant charge collection from the electron clouds in the SDD sensor layer: $S_2=A_2\cdot exp(-t/\tau_t) (1-exp(-t/\tau_f))$. The leading edge front time $\tau_f$=0.10~$\mu$s was determined by the integration time of the RC-CR filter of the pre-amplifier in the SDD output \cite{Knoll2010}, which forms the impulse response of the KETEK spectrometer. The $S_2$ form is very close to the measured pulse, as shown in the aforementioned figure by a red dashed-dotted curve which shows the difference between the approximated and real pulses, $S_2-S_R$.

The most important difference between these pulses happens in first 50~ns after their start. The nonlinear growth of the $S_R$ pulse is related to the finite charge collection time of electron clouds by the SDD anode and integration of the anode pulse by the reset pre-amplifier shown in \figref{fig:RC-CR filter}.

Fourier spectra of the $S_R$ pulse and its approximations are plotted in \figref{fig:stp}b. The spectrum of the first approximation, $F_1 \propto \tau_f/(1-\omega\tau_f)$, is shown by a blue dashed curve, and it can be seen that it decays slowly with frequency as $\omega^{-1}$. The second approximation, $F_2 \propto \tau_t/(1-\omega\tau_t)/(1+\tau_f/\tau_t-\omega\tau_f)$, is shown by a red dashed-dotted curve, and it provides a higher suppression of the output signals at frequencies $\omega>1/\tau_f$. The response spectrum, $F_R$, is presented by a black solid curve. It is significantly damped above $f_c\approx$8~MHz compared with the spectra of the approximations. This difference underlies in the restriction of the output pulse widths of standard shapers, as will be shown further on.

The noise spectrum in the output of the detector system is shown by a green thick dashed curve. The noise was measured as the voltage variations of the ADC output for a bandwidth of 25~MHz during a time interval of 15 ms when the detector was in darkness. The standard deviation of the electronic noise was estimated to be approximately 55~eV. The amplitude of the noise spectrum decays with $\omega^{-3}$ for frequencies above the cut-off frequency, $f_c$=8~MHz. The signal spectrum $F_R$ under the noise level in the high-frequency band was measured by averaging many response pulses.     

The measured parameters of the KETEK spectrometer impulse response were used for the design of a true Gaussian shaper of a given pulse width. 

\section{True Gaussian shaper of detector pulses }
\label{sec:Shaper}

Trapezoidal, cusp, and pseudo-Gaussian standard shapers \cite{jordanov1994digital,jordanov1994digital1} are designed for shaping pulses with an exponentially decaying tail and a very short rise time $\tau_r$, i.e. pulses of type $S_1$. The shaped pulses get an asymmetrical form when the shaping width approaches the rise time $\tau_r$ of the input pulse. The actual width of the output shaped pulses is restricted by the rise time of the input pulses, as will be shown in Sections 4 and 5.    

The amplitude spectra of input pulses with longer rise times ($\tau_r\approx T_s$) decay with $\omega^{-2}$ for frequencies above  $1/\tau_f$, as shown in \figref{fig:stp}b. The spectra of pulses with an ideal trapezoidal or cusp forms also decay above their characteristic frequency with $\omega^{-2}$. Therefore, the damped spectra of the input pulses significantly affect the shape and width of the output pulses when the inverse rise time is close to the characteristic frequency of the shaper. In other words, the output pulses get wider and more distorted when the rise time approaches the characteristic width to be provided by the shaper. Indeed, the symmetrical output pulses of standard shapers are much wider than the rise time of the input pulse, as illustrated in Section 4.

The cause of this restriction is the slow decaying spectra of ideally shaped pulses in the high-frequency band $\omega>1/\tau_f$. Short-shaped pulses are possible if their spectra are less affected by the damped spectra of the input pulses in this band. This requirement is best satisfied in true Gaussian pulses, $S_G=exp(-(t/\tau_G)^2/2)$, having a spectral amplitude with a Gaussian form: $F_G = \tau_G\cdot~exp(-(\omega\tau_G)^2/2)$.

The transmission function of the true Gaussian shaper is defined as the ratio of the Fourier transforms of the output Gaussian and input $S_R$ pulses:  

\begin{align}
  & K_G(\omega, T_s)=\frac{F_G(\omega, T_s)}{F_R(\omega)}  
\end{align}

Here, $T_s$ is the width of the Gaussian pulse at FWHM. The transmission function is calculated for the Gaussian pulse with an amplitude equal to that of the input pulse and centred on the $S_R$ time interval. Gaussian values outside this interval are replaced by zeros. This positioning of the Gaussian pulse minimizes its jumps at the start and end of the $S_R$ time interval and ensures that the transmission function satisfies the causal principle.

The impulse response $H_G(t, T_s)$ of the true Gaussian shaper for its implementation as a FIR filter is found as the inverse Fourier transform of the transmission function $K_G(\omega, T_s)$.

\begin{figure*}[h]
	\centering
	\includegraphics[width=\linewidth]{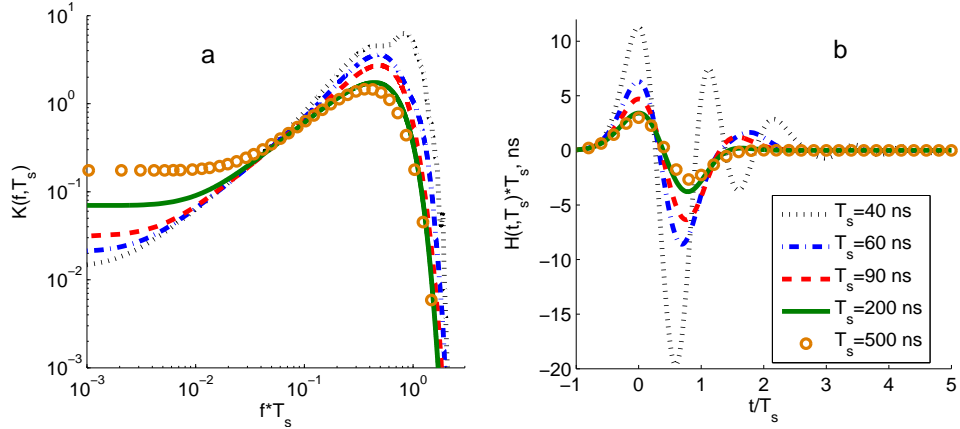}
	\caption{Transmission (a) and impulse  response (b) of true Gaussian shapers at different pulse widths}
	\label{fig:Responses}
\end{figure*}

The transmission and impulse response functions of the true Gaussian shaper calculated for the $S_R$ pulse are plotted in \figref{fig:Responses} versus frequency and time, normalized to the FWHM of the shaped pulse $T_s$. The impulse responses were multiplied by the pulse widths $T_s$ to present the plots in a closer scale.

A high gain of $K_G(\omega, T_s)$ around $f\approx1/T_s$ can be accounted for by the difference between the rise and peaking times of the input and output pulses. The short peaking times of Gaussian pulses require a high amplification in the shaper around the characteristic frequencies $f\approx1/T_s$  of these pulses. Output noise is also added in this band, and when the noise level approaches the pulse's amplitude, the true Gaussian shaping fails.

The experimental detector signal $S_D(t)$ of length $N_D$ is a sum of the responses on photons randomly arriving in time. This signal is converted to a sequence of Gaussian pulses $S_G(t)$ using the inverse Fourier transform of the product of the signal spectrum and the transmission function or by convolving `the impulse response $H_G(t, T_s)$ and the signal $S_D(t)$ digitized at a sampling period of $\Delta t$:

\begin{align}
  &  S_G(n)=\sum^{P}_{k=0}H_G(k,T_s)S_D(n-k),   ~~~~      t=n\Delta t
\end{align}

The shaper transmission function $K_G(\omega,T_s)$ must have the same length as the spectrum of the signal. For that matter, the impulse response $S_R$ is expanded to a length of $N_D$ using an exponential approximation of its tail. 

The shaping procedure uses three fast Fourier transforms, whose total computational complexity is proportional to $N_F=3\cdot N_D\cdot log_2(N_D)$. This complexity is reduced by one third for repeating measurement intervals of the same length.

The impulse response function $H_G(t, T_s)$ of the shaper or its FIR filter kernel can be truncated to a time region of several times the width $T_s$ around its maximum (see \figref{fig:Responses}b).

The number of non-zero terms P of the function is determined by the sampling frequency of the input signal, the Gaussian width $T_s$, and the required conversion accuracy. About 30 non-zero terms are needed to convert input signals digitized at a sampling rate of 50~MHz to Gaussian pulses of 90~ns FWHM with an accuracy of 0.1 \% of their maximum values. The computational complexity of this convolution is $\propto N_D\cdot P$. Thus, the computational complexity of the FIR filter and the inverse transform are equal when the signal length is approximately $2^{P/3}$ samples. This estimation is in accordance with \cite{smith2003digital}. 

True Gaussian shaping in real time requires digital signal processors (DSPs) of moderate computing power, e.g. a DSP operating at a frequency of hundreds of MHz is able to convolve an $S_R$ pulse digitized at a sampling rate of 50~MHz to a Gaussian form with the use of a 30-term kernel. The optimal convolution to a true Gaussian shape is a special issue, which should be addressed in another paper. In this work, the inverse Fourier transform was employed for accurate analysis of the true Gaussian shaper with actual recorded and simulated signals of the detection system. 

\section{Forms of the shaped pulses}
\label{sec:Forms}

\begin{figure*}[h]
	\centering
	\includegraphics[width=\linewidth]{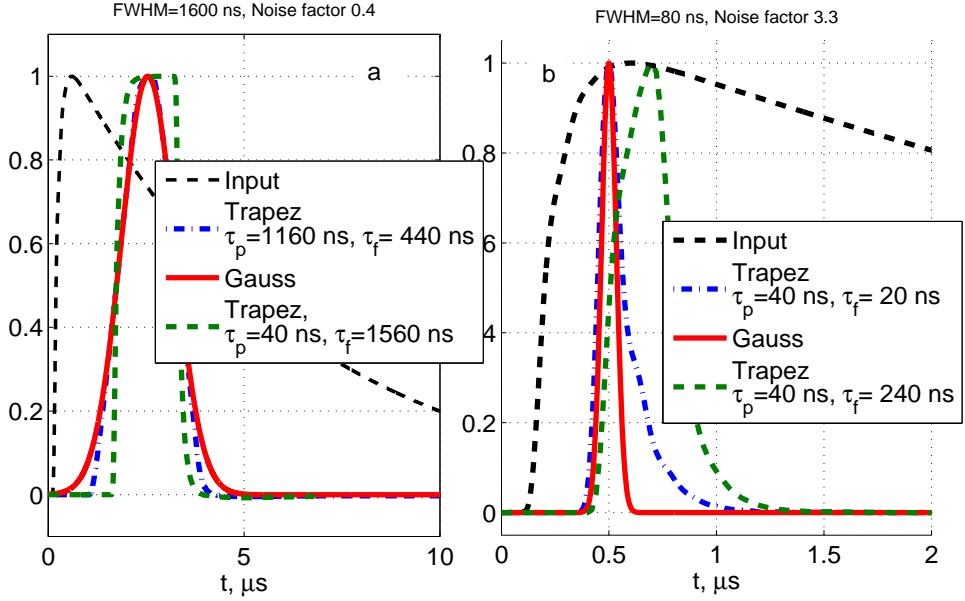}
	\caption{Normalized output pulses of true Gaussian and trapezoidal shapers\\
		a. Shaping width 1.6 $\mu$s,\\ b. Shaping width 0.08 $\mu$s}
	\label{fig:Forms}
\end{figure*}

Output pulses of the true Gaussian and trapezoidal shapers are presented in \figref{fig:Forms}a and b for the response pulses of the AXAS-D pre-amplifier, plotted in the figure by black dashed curves. The trapezoidal output was calculated with a recursive algorithm which converts an exponential pulse into a symmetrical trapezoidal form \cite{jordanov1994digital1}. The output signal of the shaper was formed as subtractions of the properly delayed and scaled digitized input signal. The shaper converts a signal with an infinite exponential tail into a symmetrical trapezoidal pulse with gain M, determined by the time constant of the input pulse tail and the sampling period; see (5) in \cite{jordanov1994digital1} for more information. The trapezoidal pulse was normalized so that its amplitude equalled that of the input $S_R$ pulse. The conversion of $S_R$ pulses with an undershoot and not exactly an exponential tail results in an asymmetrical output trapezoidal pulse which is followed by an undershot decaying tail. One can improve the form of the output tail by changing the gain coefficient M in the algorithm (see, for example, \cite{Dey2014}) but the tail cannot be completely eliminated. Further in the paper, the trapezoidal algorithm is applied with the gain value calculated for the exponential tail of the $S_1$ approximation of the detector impulse response.    

Both shapers can provide output pulses of the same width if the width of their outputs is larger than the rise time of the input pulses. Output pulses of the shapers at $\tau_{FWHM}$=1.6~$\mu$s are plotted in \figref{fig:Forms}a. The Gaussian pulse is plotted in a red solid curve. Two trapezoidal pulses of different forms and the same FWHM are plotted in blue dashed-dotted and green dashed curves. The first form is a Gaussian-like one provided by a trapezoidal shaper which time parameters are set so that to equate the width and slopes of the trapezoidal pulse to those of the Gaussian pulse measured at half maximum. The peaking and flat top times of this shaper are as follow: 

\begin{align}
  & \tau_{Tp} = \frac{\tau_{FWHM}}{2ln(2)}, ~~~~~\tau_{Tt}= \tau_{FWHM}-\tau_{Tp}\approx 0.28\tau_{FWHM} 
\end{align}

The second trapezoidal pulse is set to a rectangular-like form with short 40-ns fronts and a long 1560-ns flat top and the same FWHM. The actual output pulse has curved leading and falling edges which are considerably distorted from a pure trapezoidal form. The output noise of this shaper was three times higher than the noise levels of the two previous ones. 

A Gaussian output pulse of 0.08~$\mu$s at FWHM is shown in \figref{fig:Forms}b by a red solid curve. A Gaussian-like trapezoidal pulse at the same noise level is shown in a blue dashed-dotted curve. The parameters of this shaper  set in accordance with (3) are $\tau_{Tp}$ =20 ns and $\tau_{Tt}$ =40 ns. This pulse is strongly distorted from the trapezoidal form and its FWHM is larger than that calculated from the settings. The output noise level of the shapers, to be discussed in Section~\ref{sec:Noise}, is $\approx$8 times higher than that of the shapers with wider output pulses. The asymmetrical form of the short Gaussian-like trapezoidal pulse is certainly less advantageous for the detection of overlapped pulses. A trapezoidal pulse with a flat-top setting elongated to 240~ns, shown by a green dashed curve in \figref{fig:Forms}b, also has a strongly distorted form determined by its short 40-ns peaking time.

The output noise of the Gaussian-like trapezoidal shaper is much less than the noise of rectangular-like trapezoidal shaper because of its longer peaking time. The noise is close to the minimal level provided by trapezoidal shapers at the given pulse width.  Therefore, all further comparisons will be between true Gaussian and Gaussian-like trapezoidal shapers. Both shapers are solely characterized by the width of the output pulse at FWHM which is the sum of peaking and flat top times of an ideal trapezoidal pulse.

Trapezoidal pulses converted from the $S_R$ pulse have a long tail with an undershoot of $\approx$0.5 \% of the pulse amplitude caused by a ~0.2 \% undershoot of the $S_R$ input. This undershoot tail does not affect pulse amplitude measurements at low count rates, but should be taken into account for high count rate operation of the spectrometer, as discussed in Section~\ref{sec:Detection}.

\section{Noise factor of shapers}
\label{sec:Noise} 

Amplification of noise in a shaper is determined by its transmission function. The noise factor of shapers is defined as the ratio of the standard deviations of the output and input electronic noise measured in volts when the spectrometer is in darkness. The voltage standard deviation is expressed in photon energy by using a scale factor determined during calibration. The standard deviation of the output noise of the KETEK spectrometer, as noted earlier, is 55~eV. The spectrum of this noise compared with the spectrum of the impulse response to a photon with 5895~eV is shown in \figref{fig:stp}b. 

The noise spectrum of output shaped pulses is the transmission function of the shaper (1)  times the input noise spectrum. 

\begin{align}
& Noise_{Out}(\omega, T_s)=K_G(\omega, T_s)\cdot Noise_{In}(\omega)  
\end{align}

Thus, the noise factor of the the shaper explicitly depends on the width of the shaped pulse $T_SR$. 

The output noise of the Gaussian shaper was calculated as the inverse Fourier transform of the output noise spectrum (4). The output noise of the trapezoidal shapers is calculated from the same input noise with the use of the recursive algorithm  \cite{jordanov1994digital1} applied to generate trapezoidal response of a given pulse width from the input signal.

\begin{figure}[h]
	\centering
	\includegraphics[width=\linewidth]{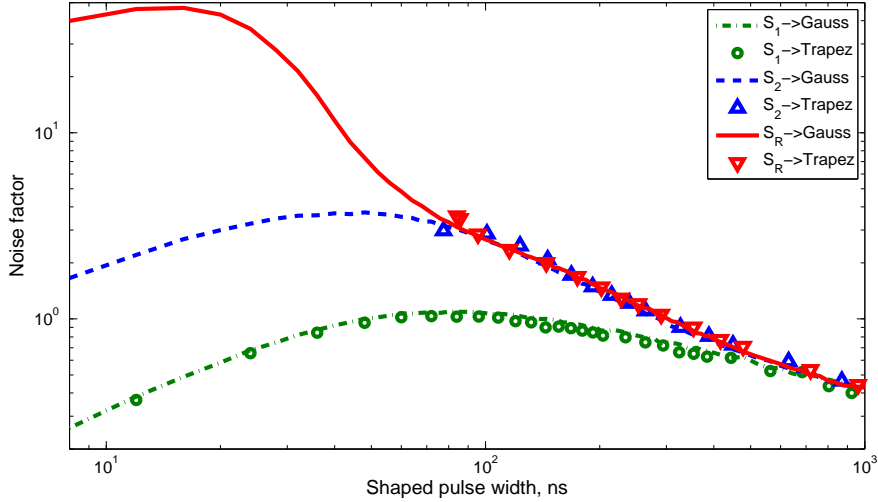}
	\caption{Noise factors of true Gaussian and Gaussian-like trapezoidal shapers. The shaped pulse width is given in terms of FWHM.}
	\label{fig:Noises}
\end{figure}

The noise factors of the shapers for the $S_R$ pulse and its two approximations are plotted in \figref{fig:Noises} versus the FWHM of the shaped pulses. The FWHM of distorted trapezoidal pulses were measured but not calculated from the set peaking and flat top times.  

True Gaussian and Gaussian-like trapezoidal shapers exhibit the same noise factors for the exponential input pulse $S_1$. These noise factors are shown in the figure by green dashed-dotted curves without and with circle marks, respectively. At large pulse widths the factors decrease inversely with the square root of the peaking time of the output pulses. The roll-off of the noise factor at small widths can be accounted for less amplification of the shapers at low frequencies and noise damping at high frequencies, i.e. noise decreases when the noise dumping frequency $f_c=\omega_c/2\pi$ becomes less than the frequency $f\approx1/T_s$ corresponding to maximum of the transmission function (see \figref{fig:stp}b and \figref{fig:Responses}a). 

Both shapers have the same noise factors if the FWHM of the shaped pulses is longer than the front time $\tau_f\approx$100~ns of the input pulses of $S_2$. These factors are plotted for the true Gaussian and trapezoidal shapers by dashed blue curves without and with open triangle marks, respectively. The shortest strongly asymmetrical output pulse of the trapezoidal shaper had a width of $\tau_f\approx$100 ns at FWHM. True Gaussian shaping allows for much shorter symmetrical output pulses, as can be seen from the red solid and blue dashed curves in \figref{fig:Noises}. 

The width of the trapezoidal output pulses converted from the $S_R$ pulses is also limited to 100~ns at FWHM as for the $S_2$ input pulses, but their form is even more distorted. True Gaussian shapers provide shorter symmetrical pulses, although at a higher noise factor, as shown by the red solid curve in \figref{fig:Noises}. The noise factor of the trapezoidal shaper is plotted by red bottom-up triangles.  

Thus, the true Gaussian shaper provides symmetrical output pulses which are much shorter than the rise time of the input pulse and, therefore, more advantageous for the detection of strongly overlapped input pulses. The noise factor of the shaper depends strongly on the form of the input pulse and the noise spectrum of the detector, which can restrict possible settings of the shaper. The detection capabilities of true Gaussian shapers are analysed in the following sections of this paper.

\section{Dead time of true Gaussian shapers}
\label{sec:DeadTime}

The detection of closely overlapped photons and the maximal output count rate of the spectrometer is determined by its dead time \cite{Knoll2010, AMPTEKGloss, AMPTEKSDD,  AMPTEKHigh}, i.e. the time interval after registration of a photon during which the system is not able to register any subsequent photons. The dead time of a trapezoidal shaper with an output pulse of undisturbed form and without pile up rejection is the sum of the peaking and flat top times of the pulse, see (2) in \cite{AMPTEKSDD}. However, this standard dead time is not certain for highly distorted short trapezoidal pulses with non-linear fronts. The peaking time is not directly applicable to the calculation of the dead time of true Gaussian shapers either. Therefore, the dead times of true Gaussian and Gaussian-like trapezoidal shapers were found in this work by means of a numerical model while taking into account the output noise of the shapers.   

The numerical model considers the impulse response of the KETEK system on two subsequent photons of similar energy and a time lag T. The detector signals are digitized at a sampling period $\tau_s$ and converted by true Gaussian or trapezoidal shapers to shorter pulses. The dead time of the shapers is found, in the proposed model, as the smallest time interval between two subsequent output pulses when: 

\begin{enumerate}[(1)]
	\item 	the sum of two single pulses has two maxima corresponding to the pulse peaks, separated by a valley; 
	\item 	the heights of both maxima compared with the valley are larger than twice the standard deviation of electronic noise at the shaper output. 
\end{enumerate}

The statistical errors of the pulse amplitude measurements are determined by the charge statistics of pair creation in the detector \cite{lechner1996pair,Schlosser2010} and electronic noises. The input electronic noises of the shapers are modelled as normal white noise smoothed by a third-order low-pass filter to fit the experimental noise spectrum shown in \figref{fig:stp}b. The standard deviation of the smoothed electronic noises was 55~eV. Instrumental or biasing errors are defined as the differences between the mean inferred peaks of two overlapped shaped pulses and their actual mean amplitudes.

These parameters were calculated for a pair of pulses with mean amplitudes of 5895~eV each versus the output pulse widths of the true Gaussian and trapezoidal shapers. Input signals were digitized at a sampling frequency of 50~MHz.

\begin{figure}[H]
	\centering
	\includegraphics[width=\linewidth]{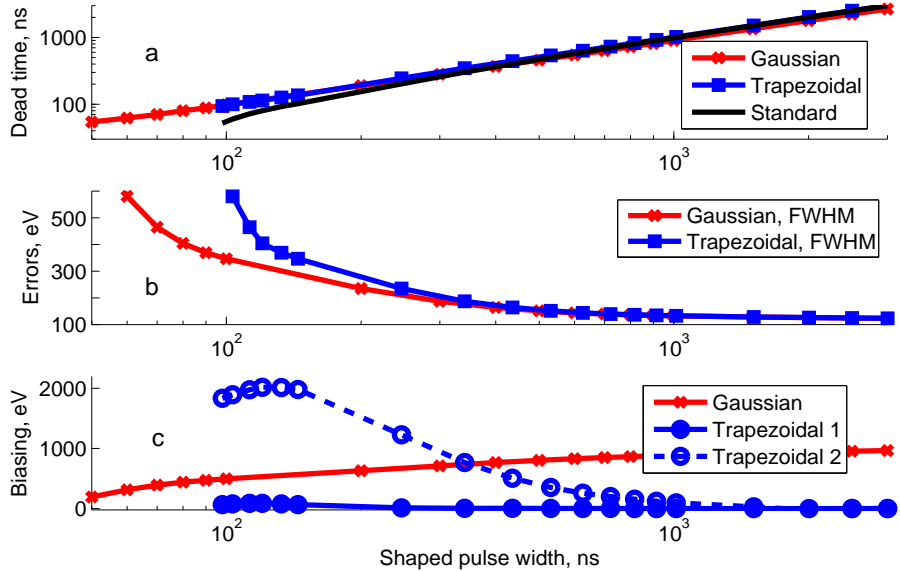}
	\caption{Dead times (a), measurement errors of single peaks (b) and biasing errors (c) of two peaks lagged by the dead time interval from each other for true Gaussian and Gaussian-like trapezoidal shapers. The widths of shaped pulses are given in terms of FWHM.}
	\label{fig:DeadTime}
\end{figure}

The dead times of the shapers, shown in \figref{fig:DeadTime}a, were nearly equal to the durations of the shaped pulses at FWHM. Both shapers had their lowest dead time determined by the sampling frequency and pulse amplitude for given detector parameters. The minimal dead time of the true Gaussian shaper was approximately half of that of the trapezoidal shapers: 60~ns versus 100~ns, respectively. The set peaking time of the trapezoidal shapers is shown in the plot by a black solid curve. Note, that short asymmetrical trapezoidal pulses are characterized by a single dead time value in accordance with the definition given in the beginning of this Section.   

The measurement errors of the pulse amplitudes were estimated as the quadratic sum of the output electronic noise of the shapers and the Fano statistics of electron hole pairs created by photon impacts in the detector area. The peak spread caused by Fano noises, estimated in \cite{lechner1996pair, Schlosser2010}, were 118~eV at FWHM. Both shapers had the same measurement errors for the peak amplitudes, and an energy resolution of $\approx$150~eV at FWHM, measured at the large shaping widths, as shown in \figref{fig:DeadTime}b. This noise level corresponds to the experimental value inferred from the experimental data for large shaping widths. The electronic noises at the shaper outputs start to dominate over the Fano noises for small shaping widths. The true Gaussian shaper performed better in terms of measurement error when the shaped pulses were shorter than the rising time $\tau_r \approx$ 200~ns of the impulse response~$S_R$.

The inferred amplitudes of two close pulses may be biased from their actual values. The ideal trapezoidal shaper is known as a biasing-free filter for large pulse widths, as confirmed in \figref{fig:DeadTime}c. Distortions of short trapezoidal pulses result in considerable biasing of the mean amplitude of the pair. A blue dashed curve with open circles in \figref{fig:DeadTime}c represents the biasing amplitudes of the second pulse in a pair, separated from the first peak by the dead time interval. On the other hand, the amplitude of the first pulse remains unbiased, as depicted in the plot by a blue solid curve with solid circles. This biasing limits the application of trapezoidal shapers with KETEK-like response pulses for dead times larger than $\approx$1~$\mu$s. 

On the contrary, true Gaussian shaping results in lower and equal biasing amplitudes for two short pulses separated by the dead time, as shown in the plot by the red curve with cross marks. For larger widths, the inferred amplitudes of two closely overlapped Gaussian pulses become strongly biased. 

The advantageous capabilities of the true Gaussian shapers of detection of very close pulses allows a considerable increase of the maximal output count rate of pulse processing, as will be shown in the next sections. Very shot true Gaussian pulses can be employed for pulse detection at poor amplitude resolution in the fast channel of the spectrometer \cite{AMPTEKSDD,  AMPTEKHigh}. The found dead time interval is used for rejection too close pulses. The remained pulses are used for the pile-up of true Gaussian pulses measured in slow channels at a higher amplitude resolution. 

\section{Resolving time of true Gaussian shapers}
\label{sec:ResolveTime}

Accurate measurements of the pulse amplitudes of both trapezoidal and Gaussian pulses in terms of amplitude error and bias require a time lag between the pair somewhat larger than the dead time. We hereby define as resolving time the smallest time interval between two detected subsequent pulses when  
\begin{enumerate}[(3)]
\item the inferred amplitudes of the overlapped detected pulses differ from the actual amplitude less than twice the standard deviation of the output electronic noise. 
\end{enumerate}

\begin{figure}[h]
	\centering
	\includegraphics[width=\linewidth]{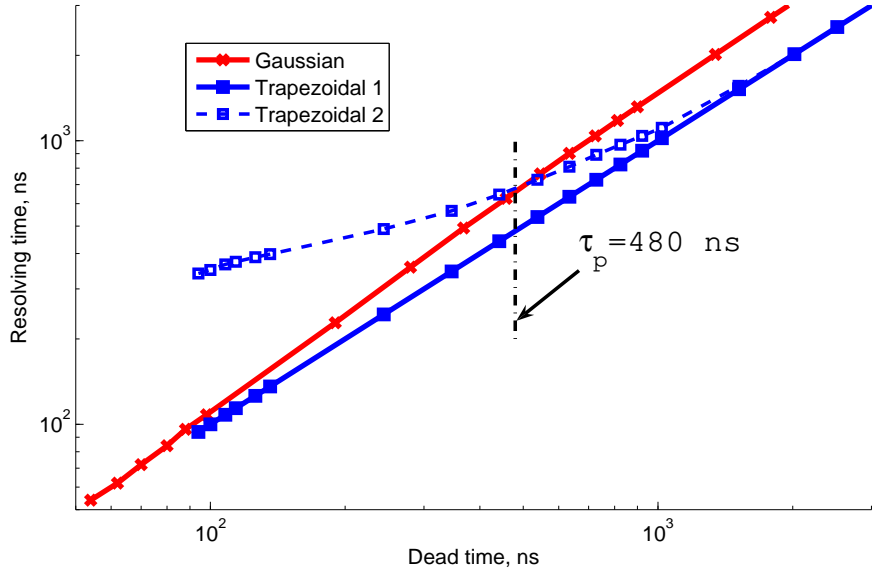}
	\caption{Resolving time of true Gaussian and Gaussian-like trapezoidal shapers.}
	\label{fig:ResolveTime}
\end{figure}

A true Gaussian pulse is resolved correctly when its time separation to both the preceding and successive pulses is larger than the resolving time. The resolving time of the true Gaussian shapers is presented in \figref{fig:ResolveTime} by a red solid curve versus their dead time.  The dead times coincide with  resolving times for very short Gaussian pulses and become larger as the pulse width is increased.

On the contrary, wide trapezoidal pulses have equal resolving and dead times. The resolving time of shorter trapezoidal pulses is split in two values. The first one relates to the first pulse of the pair and is labeled in the figure as 'Trapezoidal 1'. Its amplitude is less affected by the successive pulse and therefore the successive resolving time remains close to the dead time of the pair. The second time relates to resolution of the second pulse of the pair. Its amplitude can be strongly affected by a long tail of the preceding pulse. Therefore, the resolving time of the second pulse increases with shortening the dead time as shown in the figure by a blue dash curve.

Thus, trapezoidal shapers are better for shaped pulses of long or moderate widths, whereas Gaussian shapers are advantageous when operating with very short shaped pulses. The minimal resolving time of trapezoidal shapers is restricted by about the rise time of the $S_R$ pulse ($\tau_p \approx$480~ns), defined as the time interval between its maximum level and 1 \% of it (see \figref{fig:ResolveTime}). True Gaussian shaping reduces this limit below 100~ns and significantly increases the output count rate of the KETEK spectrometer.  

\section{Detection of pulses at a high count rate}
\label{sec:Detection}

The true Gaussian and trapezoidal shapers should be compared at high count rate operation, but the available radiation source $^{55}Fe$ could not provide a photon flux of the required intensity of $\approx 10^7$ photons per second on the detector area. Therefore, detector signals at high photon fluxes were simulated using the characteristics of the KETEK system measured at lower count rates of $\approx 10^4$ 1/s.

A radiation source was simulated to emit photons of 5895~eV energy, distributed in time according to the Poisson law. The inverse mean time interval between subsequent photons which impact the SDD sensor is the input count rate of the SDD. The output count rate of a detection system is defined as the inverse mean time interval between subsequent resolved impulse responses. 

The energy resolution of the shaped pulses is determined by the pair creation statistics in the detector layer and the electronic noise \cite{lechner1996pair} (see Sections 6 and 7). The measured standard deviation of the electronic noise of the KETEK system amounted to 55~eV. The electronic noises of the detection system were modelled as white noise smoothed by a third-order low-pass filter with a cut-off frequency of 8~MHz. The spectrum of the modelled noises corresponded to that shown in \figref{fig:stp}b. 

The simulated impulse responses at the output of the KETEK system were synchronised with impact photons. The response amplitudes are normally distributed around the mean photon energy with the standard deviation of the Fano noises \cite{Schlosser2010}. The sum of the simulated impulse responses and noises were converted by the true Gaussian and trapezoidal shapers to a sequence of shorter pulses in the way described above. The shaped signals were analysed using a numerical code which returns the times and amplitudes of the peaks detected above the threshold level, defined as five standard deviations of the output shaper noise. 

\begin{figure*}
	\centering
	\includegraphics[width=\linewidth]{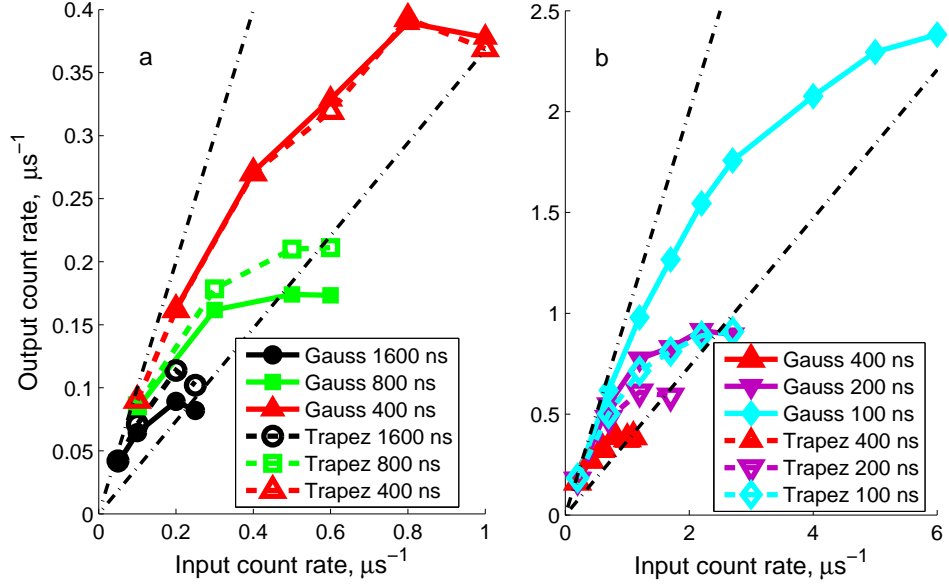}
	\caption{Output count rate with true Gaussian and Gaussian-like trapezoidal shapers for large (a) and small (b) shaping widths.}
	\label{fig:CountRates}
\end{figure*}

A true Gaussian shaper with output pulses of 60 ns  was served as a fast channel for pile-up selection of peaks detected in slow channels of the Gaussian and trapezoidal shapers, as described in \cite{Goulding1983,AMPTEKHigh}. 

The peaks detected in the slow channels were tested one by one for the time intervals to their closest preceding and successive neighbors found in the fast channel. The peaks separated less than the resolving time were removed from the counted set. The preceding and successive resolving times were applied for the pile-up selection of trapezoidal pulses. The peaks selected in the counted set will be hereafter referred to as the resolved peaks. The output count rate, energy resolution and biasing were calculated for 1000 input pulses. Therefore, the simulated data have some spread about their mean values.   

The count rates of the resolved peaks for small and large shaping widths are shown in two plots in \figref{fig:CountRates} for the true Gaussian and trapezoidal shapers. The output rates fit well to the theoretical expression $C_{out}=C_{in}\cdot exp(-2C_{in}\cdot \tau_R)$ \cite{Knoll2010, AMPTEKSDD}, where $C_{out}$ and $C_{in}$ are the output and input count rates, respectively, and $\tau_R$ is the resolving time of the system. The output rates that are equal to the input rates are plotted in the figure by the upper dashed-dotted straight line. The maximal output count rates $C_{in}/e$ achieved for $C_{in}^{max}=1/2\tau_R$ are plotted by the lower dashed-dotted straight lines. The calculated output count rates are presented only within these limits.

\begin{figure}[!h]
	\centering
	\includegraphics[width=\linewidth]{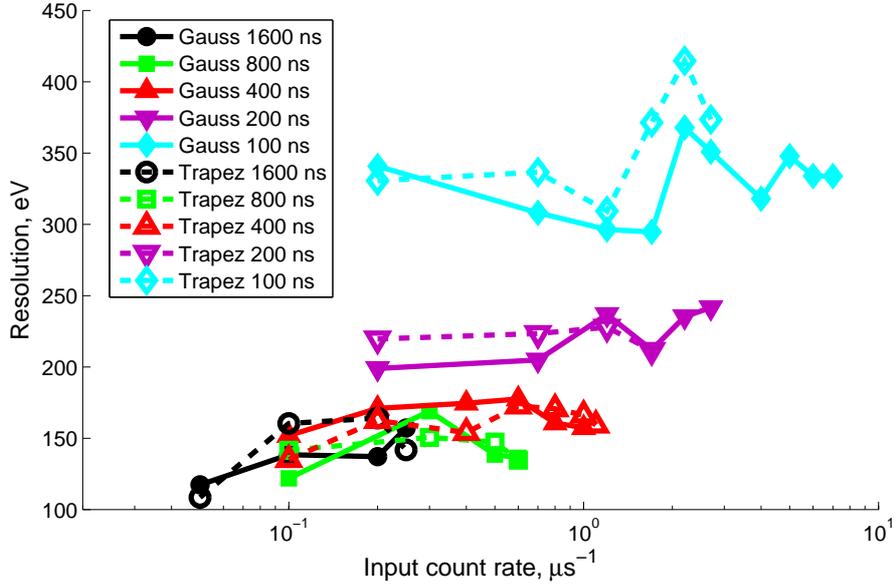}
	\caption{Amplitude resolution of pulses resolved with true Gaussian and  Gaussian-like trapezoidal shapers}
	\label{fig:Resolution}
\end{figure}

The first plot in the figure presents data for shaping widths nearly equal or larger than the rise time of the $S_R$ pulse. The output count rates of the true Gaussian shapers are lower than those of the trapezoidal shaper, as expected from~\figref{fig:ResolveTime}. 

\begin{figure}[h]
	\centering
	\includegraphics[width=\linewidth]{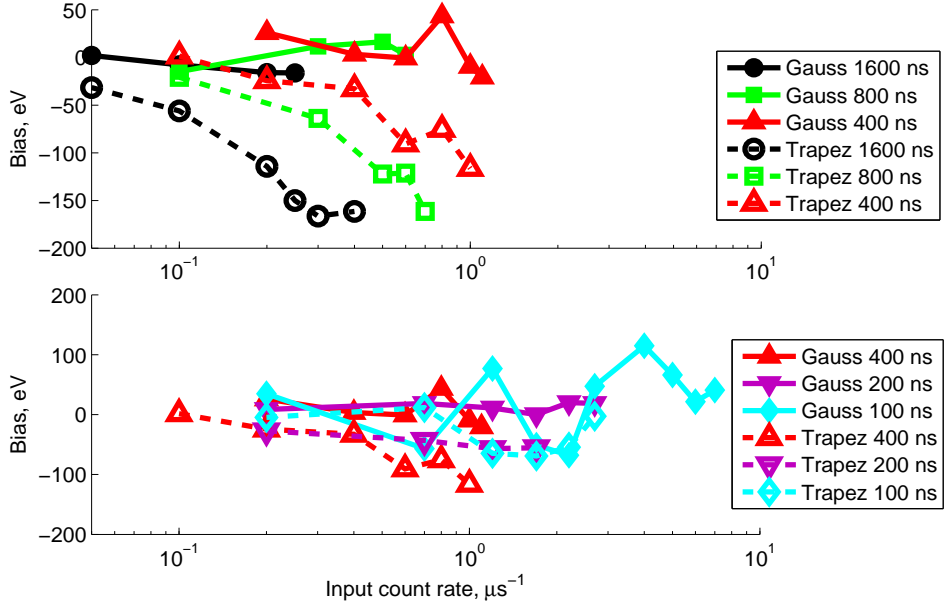}
	\caption{Biasing errors of pulse amplitude measurements with true Gaussian and  Gaussian-like trapezoidal shapers}
	\label{fig:Bias}
\end{figure}

The second plot in the figure presents the count rates calculated for small shaping widths, where true Gaussian shapers are superior and provide a lower resolving time. At small widths, the maximal output count rate of the Gaussian shapers exceeds $3\cdot10^6$~counts per second, which is more than three times higher than that calculated for the trapezoidal shapers and ten times higher the maximal rate specified by the manufacturer \cite{KETEKVITUS}. The highest output count rate, $>4\cdot10^6$~counts per second, was achieved with Gaussian pulses of 70~ns at FWHM. Obtaining higher rates is impossible because the shaping noises increase as the pulse widths decrease.

The spectral resolution of the resolved pulse amplitudes is shown in \figref{fig:Resolution}. The data are also restricted by the maximal output count rate and the curves are marked in the same way as in \figref{fig:CountRates}. The resolution is expressed in terms of the FWHM of the spectral peak around the mean photon energy of 5895~eV.
 
The resolutions of the both types of shapers providing long shaped pulses are in the range of the technical specifications of the tested system. At higher rates the resolution of the true Gaussian and trapezoidal shapers of the same pulse widths are similar for count rates lower than the maximal output rate of the trapezoidal shapers. The calculated resolution and output count rate of all trapezoidal shapers presented in the Section correspond well to those of a fast SDD spectrometer developed by AMPTEK \cite{AMPTEKHigh}. 

The true Gaussian shapers surpass significantly the maximal output count rate of the trapezoidal shapers of the same pulse width without losses of energy resolution. 

\begin{figure}[h]
	\centering
	\includegraphics[width=\linewidth]{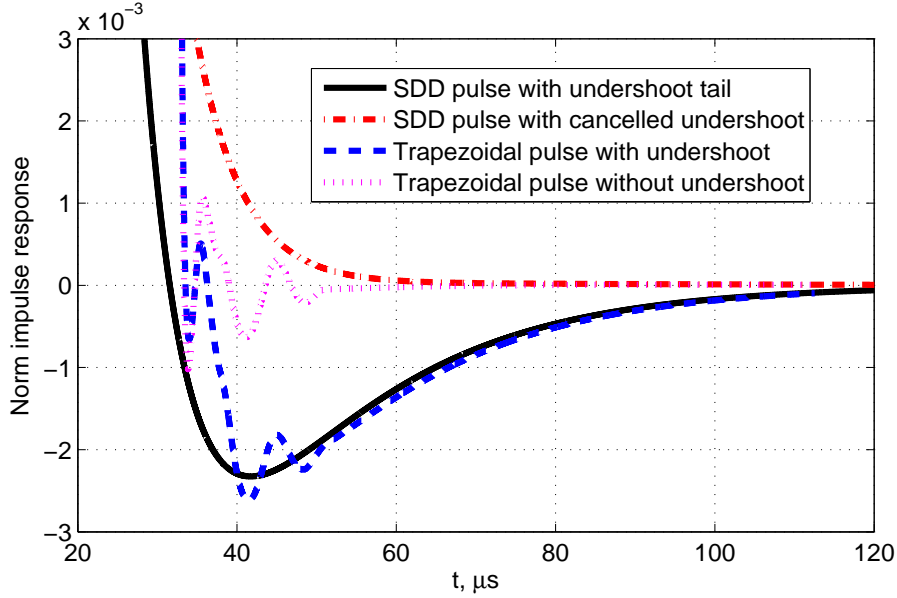}
	\caption{Tails of the KETEK and shaped pulses with and without digital pole-zero cancellation}
	\label{fig:UndershootCancellation}
\end{figure}   

The inferred amplitudes can be biased from the mean amplitude of the input pulses. The biased amplitudes of the shaped signals for the true Gaussian and trapezoidal shapers are shown in \figref{fig:Bias}. True Gaussian shaping provides a much lower bias of the resolved pules than trapezoidal shaping. 

The large negative biasing amplitudes of trapezoidal pulses are caused by the accumulation of long undershoot tails of many pulses. These undershoots come from a 0.2 \% undershoot of the $S_R$ pulse, as shown in \figref{fig:stp}a. The undershoot of this pulse is not perfectly cancelled with the pole-zero circuit of the preamplifier \cite{Knoll2010}. The further cancellation was done with a digital reversed pole-zero circuit which was found to significantly reduce the tail of the KETEK spectrometer impulse response, as shown in \figref{fig:UndershootCancellation}. The measured undershoot tail is shown by a solid black curve. The output tail after the digital cancellation filter is plotted by a dashed-dotted red curve. The tails of the trapezoidal pulses converted from the measured and filtered $S_R$ pulses are plotted in dashed blue and dotted magenta curves, respectively. The cancellation filter allows for a reduction of biasing in the trapezoidal amplitudes down to the biasing of Gaussian pulses, as shown in \figref{fig:Bias}, and does not change other characteristics of the trapezoidal shapers. Note that the pole-zero cancellation does not affect the biasing in closely overlapped trapezoidal pulses considered in Section \ref{sec:DeadTime} and shown in \figref{fig:DeadTime}c.

\section{Variations in the amplitudes of shaped pulses}
\label{sec:Variations}

An impact photon creates an electron cloud in the sensor layer of the SDD. The cloud drifts in the electric field to the SDD anode, expending in size. Therefore, the charge collection time of the cloud depends on the distance between the anode and the impact point \cite{Gatti1987, Prigozhin2012}. Thus, distribution of collected photons across a large detector area can result in the ballistic deficit of shaped pulses, i.e. variations of their amplitudes with the charge collection time. 

\begin{figure}[h]
	\centering
	\includegraphics[width=\linewidth]{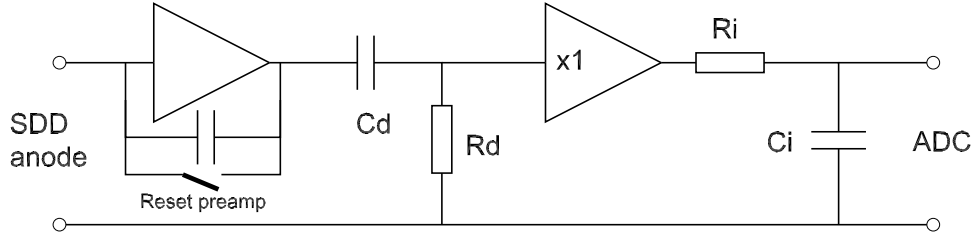}
	\caption{Reset pre-amplifier and RC-CR filter of SDD anode signals}
	\label{fig:RC-CR filter}
\end{figure}

The amplitude variations of the shaped pulses were analysed by means of the following model. First, the charge collection time is estimated with the use of \cite{Gatti1987} for the KETEK H7 SDD in terms of the rms widths of the anode current pulses $\tau_C$. These pulses are integrated by a reset preamplifier, as shown in \figref{fig:RC-CR filter}, which converts them to step-wise voltage signals. The rise time of these signals is determined by the charge collection time, and their amplitude is proportional to the photon energy used. The step-wise signals are shaped by means of an RC-CR filter connected to the output of the reset preamplifier, as shown in \figref{fig:RC-CR filter}. 

\begin{figure}[h]
	\centering
	\includegraphics[width=\linewidth]{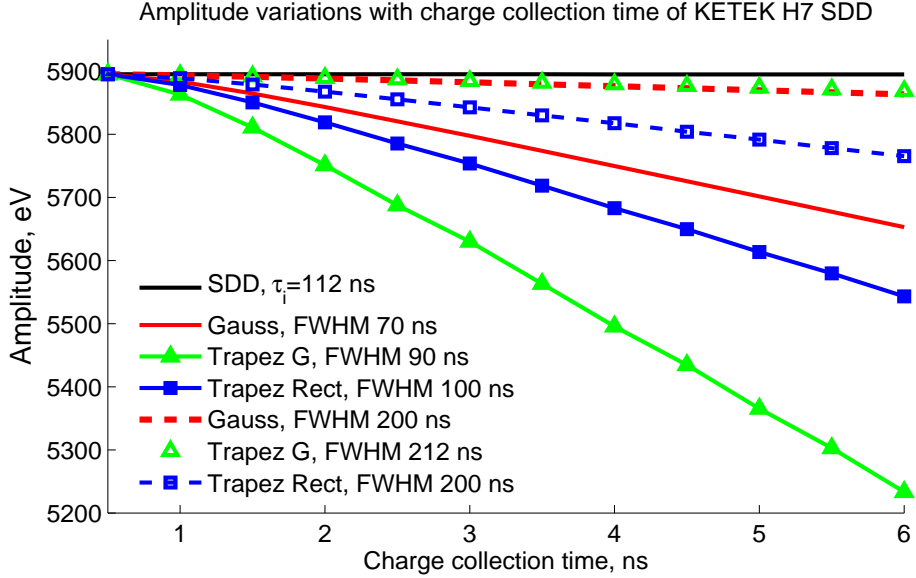}
	\caption{Amplitude variations of shaped pulses of the KETEK spectrometer}
	\label{fig:AmplVariationsH7}
\end{figure}

The maximal rms width of the current pulses of the KETEK H7 SDD found from \cite{Gatti1987} is 6~ns. The integration time of the filter $\tau_i =R_iC_i$ determines the rise time of the impulse response, whereas its decay is specified by the differentiation time $\tau_d =R_dC_d$. The measured impulse response $S_R$ of a small area of the H7 SDD was fitted using the model for $\tau_i$ =112~ns and $\tau_d$ =5945~ns.

The output amplitudes of a true Gaussian shaper and two trapezoidal shapers were calculated using the model and the suggested parameters of the KETEK spectrometer. A Gaussian shaper of width $\tau_{G0}$ was designed and calibrated to the photon energy at the lowest charge collection time of the detector, and was applied for collection times in the full range. The first trapezoidal shaper converted the SDD pulses into a Gaussian-like trapezoidal form. The second shaper was designed for trapezoidal pulses with the lowest peaking time (one sampling period set for the leading edge of the pulse) and longest flat top. The FWHM of both trapezoidal shapers were set to that of the Gaussian pulse $\tau_{G0}$.

The amplitude variations of the shaped pulses versus the charge collection time $\tau_C$ of the KETEK H7 SDD were calculated for a sampling frequency of 50~MHz and are shown in \figref{fig:AmplVariationsH7}. The amplitudes of the SDD pulses are plotted in a black curve and are independent from $\tau_C$. The amplitudes of Gaussian pulses of 70~ns at FWHM decreased by 200~eV with $\tau_C$, as shown by the red solid curve. This is one third of the spectrometer resolution calculated in the previous section for the same pulse width.

The amplitude variations of both types of trapezoidal pulses are plotted in the figure by green solid triangles and blue solid squares. They ranged from 300 to 600~eV and were larger than the calculated energy resolution of both types of trapezoidal shapers. 

\begin{figure}[h]
	\centering
	\includegraphics[width=\linewidth]{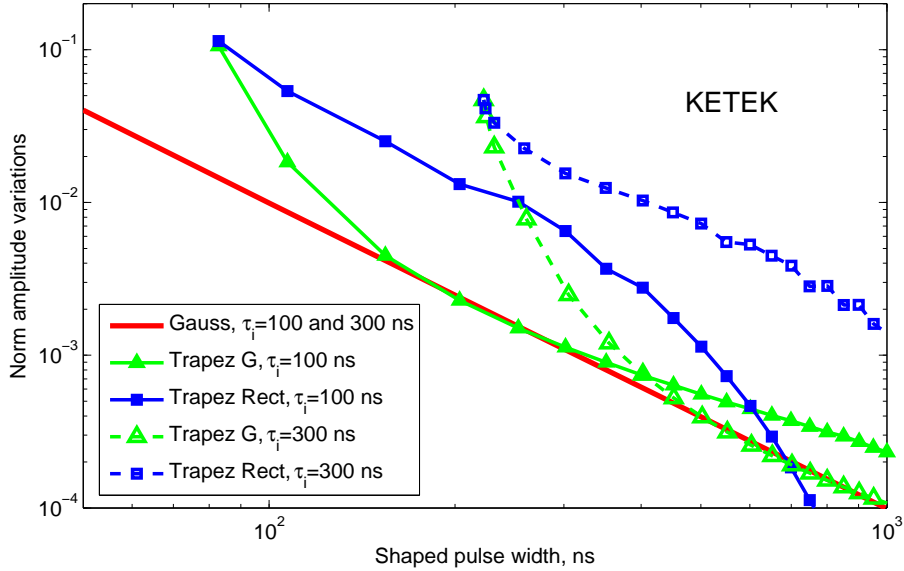}
	\caption{Normalized range of amplitude variations with shaped pulse width at fixed integration times for KETEK spectrometer. The widths of shaped pulses are given in terms of FWHM.}
	\label{fig:AmplVariationsKETEK}
\end{figure}
 
Gaussian amplitude variations decrease with shaped pulse width, and for $\tau_{G0}$=200~ns they amount to $\approx$30~eV, i.e. they are negligible with the SDD resolution at low count rates. They are plotted in the figure by a red dashed curve. The first type of trapezoidal shaper mentioned exhibited the same variations, shown in the figure by green open triangles over the red dashed curve. The second type of trapezoidal shaper with rectangular output pulses had variations several times larger, which are plotted by a blue dashed curve with open squares in the figure. 

\begin{figure}[h]
	\centering
	\includegraphics[width=\linewidth]{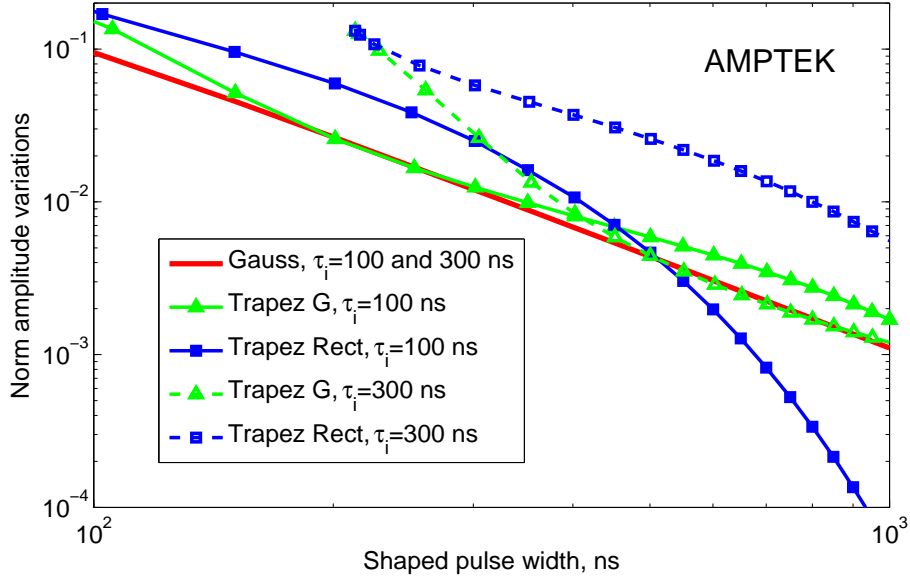}
	\caption{Normalized range of amplitude variations with shaped pulse width at fixed integration times of AMPTEK XR100 SDD. The widths of shaped pulses are given in terms of FWHM.}
	\label{fig:AmplVariationsAMPTEK}
\end{figure}

The model used shows that amplitude variations of Gaussian pulses are independent of the integration time $\tau_i$. The amplitude variations of trapezoidal pulses depend both on integration time and on pulse width, as shown in \figref{fig:AmplVariationsKETEK}. Here, the range of amplitude variations at a fixed integration time was normalized to the mean amplitude of the shaped pulses of the KETEK spectrometer. The minimal width of the trapezoidal pulses is limited to $\approx80 \%$ of the integration time $\tau_i$. The amplitude variations of short trapezoidal pulses are considerably larger than that of Gaussian pulses. The variations of the first type of trapezoidal pulses decreased with width and were close to those of Gaussian pulses, as shown by the green curves with triangles in the figure. The amplitude variations of the second type of trapezoidal pulses decreased slowly with width and remained significantly larger than that of Gaussian pulses for small and moderate pulse widths. The form of rectangular-like trapezoidal pulses is more affected by variations of the charge collection time because their peaking time is closer to the range of the charge collection time.     

Thus, the amplitude variations of short true Gaussian pulses can be neglected when compared with the energy resolution of the KETEK spectrometer. The variations of the short trapezoidal pulses are much larger and therefore the energy resolution of the spectrometers operating with short pulses at high count rates can be reduced. 

The amplitude variations might be higher in SDDs with a larger area. The variations in charge collection time in an AMPTEK XR100-SDD \cite{AMPTEKXR100}, estimated in a similar way \cite{Gatti1987}, amounted to $\approx20$~ns. This value corresponds to direct measurements carried out in \cite{Prigozhin2012}. Normalized amplitude variations calculated for this detector at integration times of 100~ns and 300~ns are shown in  \figref{fig:AmplVariationsAMPTEK}. Larger variations of the charge collection time in the XR100 SDD result in higher variations of the shaped pulse amplitudes. At short pulse widths, the variations exceed the energy resolution of the detection system. True Gaussian shapers provide the smallest variations at high count rates.

\section{Discussion}
\label{sec:Discussion}

The symmetrical form of true Gaussian pulses is one of the factors which enable a high count rate when using them. One can try to design a true trapezoidal shaper in a similar way, i.e. with the use of a transmission function found as the ratio of Fourier transforms of the output trapezoidal and input pulses. Such a trapezoidal shaper was developed for scintillator pulses with a short rising time \cite{Redus2014, RedusPrivate}.  

The spectrum of a symmetrical trapezoidal pulse is much wider than that of a Gaussian pulse of the same width and decays slowly in the high frequency band: $F_{trap}=4/(\omega^2T)~sin(\omega(\tau_p+T)/2)sin(\omega~T/2) \approx 4/(\omega^2T) $. Here, $\tau_p$ is the peaking time of the pulse and T is its flat top duration. The amplification of the shaper at high frequencies $f\ge~1/\tau_p$ must be high enough to form a true trapezoidal pulse of a width less than the rise time of the input pulse. In this case, the shaper transmission has little slope or is not even damped at high frequencies, resulting high output noise. The same conclusion holds true for cusp-like and pseudo-Gaussian shapers. They use parabolic approximations of the output pulse form, in which the Fourier spectra decays with $\omega^{-3}$. The large noise factors of standard shapers with output pulses of true forms limit their application in spectrometers at high count rates.    

\section{Conclusion}

A new digital true Gaussian shaper for the detection of pulses at high count rates has been proposed in this paper. The shaper was analysed for the impulse response of an AXAS-D system with a H7 VITUS SDD, made by KETEK GmbH, in terms of noise factor, dead and resolving times, energy resolution and output count rate of the measurements. The analysis employed numerical models based on calibrations and measured characteristics of the KETEK spectrometer. The main advantages of the proposed true Gaussian shaper compared with standard shapers are as follows: 

1. the width of symmetrical output pulses of the Gaussian shapers can be reduced well below the rise time of the input pulse to overcome the minimal width of output of standards shapers,   

2. short Gaussian pulses allow detection and accurate measurements of amplitudes of strongly overlapped input pulses, 

3. true Gaussian shapers provide a significant increase of the output count rate of SDD spectrometers while maintaining their energy resolution. 

4. Amplitude variations of Gaussian pulses with charge collection time of electron clouds by the SDD anode are less or similar to those provided by standard shapers.

\section*{Acknowledgments}
This work was supported by a grant of the Russian Science Foundation (17-12-01110) and by the Ioffe Institute.
 
\bibliography{citations}{} 

\begin{thebibliography}{10}
\expandafter\ifx\csname url\endcsname\relax
  \def\url#1{\texttt{#1}}\fi
\expandafter\ifx\csname urlprefix\endcsname\relax\def\urlprefix{URL }\fi
\expandafter\ifx\csname href\endcsname\relax
  \def\href#1#2{#2} \def\path#1{#1}\fi

\bibitem{smith2003digital}
S.~Smith, Digital signal processing : a practical guide for engineers and
  scientists, Newnes, Amsterdam Boston, 2003.

\bibitem{proakis2001digital}
J.~G. Proakis, D.~G. Manolakis, Digital signal processing: principles
  algorithms and applications, Prentice Hall, 2001.

\bibitem{jordanov1994digital}
V.~T. Jordanov, G.~F. Knoll, Digital synthesis of pulse shapes in real time for
  high resolution radiation spectroscopy, Nuclear Instruments and Methods in
  Physics Research Section A: Accelerators, Spectrometers, Detectors and
  Associated Equipment 345~(2) (1994) 337--345.

\bibitem{jordanov1994digital1}
V.~T. Jordanov, G.~F. Knoll, A.~C. Huber, J.~A. Pantazis, Digital techniques
  for real-time pulse shaping in radiation measurements, Nuclear Instruments
  and Methods in Physics Research A 353~(1) (1994) 261--264.

\bibitem{KETEKAXAS}
\href{http://www.ketek.net/products/axas/axas-d}{{AXAS-D Analytical X-ray
  Acquisition System - KETEK GmbH}}.
\newline\urlprefix\url{http://www.ketek.net/products/axas/axas-d}

\bibitem{KETEKVITUS}
\href{https://www.ketek.net/sdd/vitus-sdd-modules/vitus-h7/}{{VITUS H7 -
  Silicon Drift Detector (SDD) - KETEK GmbH}}.
\newline\urlprefix\url{https://www.ketek.net/sdd/vitus-sdd-modules/vitus-h7/}

\bibitem{Gatti1987}
E.~Gatti, A.~Longini, P.~Rehak, M.~Sanpietro, Dynamics of electrons in drift
  detectors, Nuclear Instruments and Methods in Physics Research A 253~(3)
  (1987) 393--399.

\bibitem{Metzger2004}
W.~Metzger, J.~Engdahl, W.~Rossner, O.~Boslau, J.~Kemmer, Large-area silicon
  drift detectors for new applications in nuclear medicine imaging, IEEE
  Transactions on Nuclear Science 51~(4) (2004) 1631--1635.

\bibitem{lashkul2001effect}
S.~Lashkul, V.~Budnikov, E.~Vekshina, V.~D’yachenko, V.~Ermolaev, L.~Esipov,
  E.~Its, M.~Y. Kantor, D.~Kuprienko, A.~Y. Popov, et~al., Effect of the radial
  electric field on lower hybrid plasma heating in the ft-2 tokamak, Plasma
  Physics Reports 27~(12) (2001) 1001--1010.

\bibitem{Goulding1983}
F.~S. Goulding, D.~A. Landis, N.~W. Madden,
  \href{http://ieeexplore.ieee.org/document/4332275/}{Design philosophy for
  high-resolution rate and throughput spectroscopy systems}, IEEE Transactions
  on Nuclear Science 30~(1) (1983) 301--310.
\newblock \href {http://dx.doi.org/10.1109/TNS.1983.4332275}
  {\path{doi:10.1109/TNS.1983.4332275}}.
\newline\urlprefix\url{http://ieeexplore.ieee.org/document/4332275/}

\bibitem{Knoll2010}
G.~F. Knoll, Radiation Detection and Measurement, 4th Edition, John Wiley and
  Sons, Inc., Hoboken, NJ, USA, 2010.

\bibitem{Dey2014}
M.~Dey, M.~Biswas, S.~Ghosh, S.~Chakaraborty, Real time pulse processors for
  physics experiments - simulation and implementation, in: International
  Conference on Signal Propagation and Computer Technology (ICSPCT), 2014, pp.
  726--731.

\bibitem{AMPTEKGloss}
\href{http://www.amptek.com/pdf/glossary.pdf}{{AMPTEK General tutoirials.
  Glossary}}.
\newline\urlprefix\url{http://www.amptek.com/pdf/glossary.pdf}

\bibitem{AMPTEKSDD}
\href{https://amptek.com/pdf/ansdd1.pdf}{{Amptek Silicon Drift Diode (SDD) at
  High Count Rates}}.
\newline\urlprefix\url{https://amptek.com/pdf/ansdd1.pdf}

\bibitem{AMPTEKHigh}
\href{http://amptek.com/pdf/andpp2.pdf}{{Amptek Operating the DP5 at High Count
  Rates}}.
\newline\urlprefix\url{http://amptek.com/pdf/andpp2.pdf}

\bibitem{lechner1996pair}
P.~Lechner, R.~Hartmann, H.~Soltau, L.~Str{\"u}der, Pair creation energy and
  fano factor of silicon in the energy range of soft x-rays, Nuclear
  Instruments and Methods in Physics Research A 377~(2-3) (1996) 206--208.

\bibitem{Schlosser2010}
D.~Schlosser, P.~Lechner, et.al, Expanding the detection efficiency of silicon
  drift detectors, Nuclear Instruments and Methods in Physics Research A 624
  (2010) 270--276.

\bibitem{Prigozhin2012}
G.~Prigozhin, K.~Gendreau, R.~Foster, G.~Ricker, J.~Villasenor, J.~Doty,
  S.~Kenyon, Z.~Arzoumanian, R.~Redus, A.~Huber,
  \href{http://adsabs.harvard.edu/abs/2012SPIE.8453E..18P}{Characterization of
  the silicon drift detector for nicer instrument}, in: High Energy, Optical,
  and Infrared Detectors for Astronomy V, Vol. 8453, 2012, p. 845318.
\newblock \href {http://dx.doi.org/10.1117/12.926667}
  {\path{doi:10.1117/12.926667}}.
\newline\urlprefix\url{http://adsabs.harvard.edu/abs/2012SPIE.8453E..18P}

\bibitem{AMPTEKXR100}
\href{http://www.amptek.com/wp-content/uploads/2014/04/XR-100SDD-Silicon-Drift-Detector-SDD-Specifications.pdf}{{Amptek
  XR100-SDD specifications}}.
\newline\urlprefix\url{http://www.amptek.com/wp-content/uploads/2014/04/XR-100SDD-Silicon-Drift-Detector-SDD-Specifications.pdf}

\bibitem{Redus2014}
R.~Redus, A.~Huber, et. al.,
  \href{http://nssmic2014.npss-confs.org/Images/Program/2014Program.pdf}{General
  purpose digital processors for scintillation spectroscopy}, in: 2014 IEEE
  Nuclear Science Symposium and Medical Imaging Conference. The 21st
  International Symposium on Room-Temperature Semiconductor Detectors, IEEE,
  2014, pp. N23--7.
\newline\urlprefix\url{http://nssmic2014.npss-confs.org/Images/Program/2014Program.pdf}

\bibitem{RedusPrivate}
personal communication.

\end{thebibliography}
\end{document}